\documentclass[twocolumn,preprintnumbers,superscriptaddress,nofootinbib,prd,aps,10pt]{revtex4-1}
\pdfoutput=1

\usepackage{graphics}
\usepackage[dvips]{graphicx}
\usepackage{mathrsfs}
\usepackage{amssymb}
\usepackage{amsmath}
\usepackage{verbatim}
\usepackage{float}
\usepackage{slashed}
\usepackage{bbm}
\usepackage{topcapt}
\usepackage{hyperref}

\newcommand{\be}{\begin{equation}}
\newcommand{\ee}{\end{equation}}
\newcommand{\bea}{\begin{eqnarray}}
\newcommand{\eea}{\end{eqnarray}}

\newcommand{\mev}{\text{MeV}}
\newcommand{\gev}{\text{GeV}}
\newcommand{\tev}{\text{TeV}}
\newcommand{\lum}[2]{#1 \times 10^{#2} \text{~cm}^{-2}\text{sec}^{-1}}
\newcommand{\ifb}{\text{~fb}^{-1}}

\newcommand{\ra}{\rightarrow}
\newcommand{\fb}{\text{fb}}

\begin{document}
\title{The Muon Collider as a $H/A$ factory}
\author{Estia Eichten}
\affiliation{Theoretical Physics Department, Fermilab, Batavia, IL 60510}
\author{Adam Martin}
\affiliation{PH-TH Department, CERN, CH-1211 Geneva 23, Switzerland}
\affiliation{Department of Physics, University of Notre Dame, Notre Dame, IN 46556,
             USA\,}~\thanks{visiting scholar}

\preprint{FERMILAB-PUB-13-193-T}
\preprint{CERN-PH-TH/2013-111}

\begin{abstract}
We show that a muon collider is ideally suited for the study of heavy $H/A$ scalars, cousins of the Higgs boson found in two-Higgs doublet models and required in supersymmetric models.   The key aspects of $H/A$ are: (1) they are narrow, yet have a width-to-mass ratio far larger than the expected muon collider beam-energy resolution, and (2)  the larger muon Yukawa allows efficient $s$-channel production. We study in detail a representative Natural Supersymmetry model which has a $1.5 \, \tev$ 
$H/A$ with $m_H - m_A = 10 \, \gev$.  The large event rates at resonant peak allow the determination of the individual $H$ and $A$ resonance parameters (including CP) 
and the decays into electroweakinos provides a wealth of  information unavailable to any other present or planned collider.
\end{abstract}

\maketitle

\section{Introduction}
\label{sec:intro}

 The discovery at the LHC of a standard model (SM) like Higgs at $m_H = 125\, \gev$~\cite{Aad:2012tfa,Chatrchyan:2012ufa} has greatly 
 altered our theoretical expectations.
The couplings of this state to electroweak gauge bosons makes it clear that the majority (if not all) of electroweak symmetry breaking is associated with this SM-like Higgs boson. 

Within supersymmetric models, this Higgs  mass is at the upper end of previous expectations and this, coupled with the lack of any experimental evidence (to date) for supersymmetric particles, has lead to a reevaluation of viable models.  In particular, theoretical efforts have  focussed on avoiding serious fine-tuning while still allowing a portion of the sparticle masses at scales well above a TeV.  Phenomenological efforts are focussed on finding at what scale new physics will arise and what is the best machine to probe that scale.  

It is required in supersymmetric models that there are other spin-zero states left behind after electroweak symmetry breaking. More generally, heavy neutral Higgses $H/A$ and charged Higgses $H^\pm$, of  two Higgs-doublet models (2HDM), are a simple and well-motivated possibility for new physics.  Despite their intimate connection electroweak symmetry breaking, the $H/A$ interact weakly with the Higgs boson itself, thus $H/A$ can be quite heavy without rendering these scenarios unnatural. 
In this limit of large $m_A$, the masses of all the $H, A, H^\pm$ states become nearly degenerate. Additionally, large $\tan\beta$ suppresses the couplings to up-sector fermions, and the near SM-like coupling of the 125 GeV Higgs to electroweak gauge bosons implies that  
couplings of $H,A$ and $H^{\pm}$  to $W^+W^-$ and $Z^0Z^0$ are greatly suppressed; as a result, the heavy Higgses remain relatively narrow.
As $m_A$ is increased, $H/A$ become increasingly difficult to discover at the LHC, yet they remain targets for next-generation lepton colliders. In this paper we therefore evaluate the prospects for discovery and precision studies of heavy $H/A$ at a muon collider. While we will focus on $H/A$ within supersymmetric setups (so-called Type-II 2HDM), our results apply to a wider class of 2HDM models.

As we will show, a muon collider with nominal beam resolution and luminosity parameters ($R = 1.0 \times 10^{-3}$ and luminosity $>10^{34}$ cm$^2$ sec $^{-1}$  \cite{Delahaye2013}) is the ideal collider for the study of $H/A$ properties and decay product for any $H$ and $A$ masses between the present LHC bounds and up  to at least $ 6 \, \tev $.   Although, some muon collider studies have
 been performed previously for relative low mass $H/A\, (\sim 400 \gev$)~\cite{Barger:2001mi,Dittmaier:2002nd}, these studies mainly focussed on the observability of the states as separate resonances.  In light of the present reevaluation of viable supersymmetry models we consider heavier $H/A$ -- up  to at least $ 6\, \tev$ -- and find new and exciting opportunities for $H/A$ factories.
 
The setup of this paper is the following. In Section \ref{sec:othermachines} we will briefly review the limits and $H/A$ discovery potential at the LHC and at prospective $e^+e^-$ machines. Next, in Section \ref{sec:MC}, we describe the basics of the muon collider and the s-channel resonance production of heavy neutral Higgs.  The resonance production of $H/A$ in various alternative supersymmetry scenarios are 
presented in Section \ref{sec:bench}. Although the features we are studying are general, we will focus in detail on a representative `Natural Supersymmetry' model constructed to evade the present LHC bounds and still have particles accessible to a $1\, \tev$ $e^+ e^-$\,Linear Collider.  We show the performance of  a muon collider in measuring the parameters of these $H$ and $A$ states in Section \ref{sec:NS}.   Then in Section \ref{sec:factory} we study what can be learned 
from the decays into both SM and supersymmetric particles. Finally, we conclude with a discussion of how to find 
these extra scalar states in Section \ref{sec:disc} 

\section{$H/A$ at the LHC and $e^+e^-$ colliders}
\label{sec:othermachines}

Heavy Higgses have been searched for at the LHC by both ATLAS~\cite{Aad:2011rv, atlas2,Aad:2012dpa} and CMS~\cite{CMSHA,Behr:2013ji, cms2,Chatrchyan:2012yca} using a variety of final states. Cast in term of Type-II 2HDM parameters,  the current bounds exclude $m_{H/A} \lesssim 300\,\gev$ for $\tan{\beta} = 10$, stretching to $m_{H/A} \lesssim 600\, \gev$ for $\tan{\beta} = 40$~\cite{ Baglio:2011xz, Carena:2011fc, Christensen:2012ei, Chang:2012zf, Carena:2013qia, Arbey:2013jla, Djouadi:2013vqa}. After the full 8 TeV dataset has been analyzed, these bounds are expected to increase by an additional $50-100\,\gev$, and the limits are estimated to approximately double ($m_{H/A} \lesssim 900\, \gev,\, \tan{\beta} = 10,\, \text{and}\,m_{H/A} \lesssim 1.5\, \tev, \tan{\beta} = 40 $) after $150\, \ifb$ of 14 TeV LHC running~\cite{Arbey:2013jla}. Indirect bounds from precision studies of the couplings of the $125\, \gev$ state may also give some insight~\cite{Giardino:2013bma, Maiani:2012qf, Arbey:2013jla}, but these bounds depend on assumptions about loop contributions from other light states in the spectrum.

The discovery potential of $H/A$ at TeV scale $e^+e^-$ colliders has also been studied in detail~\cite{Phinney:2007gp,Lebrun:2012hj,Miyamoto:1425915,Brau:2012hv}. At an $e^+e^-$ collider, heavy Higgses must be produced in association with some other particle, since the electron Yukawa coupling is too small for efficient $s$-channel production. One popular production mode is $e^+e^- \ra H Z^0$, however this process suffers from a small $HZ^0Z^0$ coupling, as we will explain in more detail later. Other production modes include $e^+e^- \ra t\bar t\, H/A$ and $e^+e^- \ra HA$. While these modes yield distinct final states, the cross sections are usually so low that the rate at a prospective linear collider is insufficient for precision studies. Additionally, the fact that the heavy Higgses must be produced in association makes extracting the $H/A$ masses and widths challenging.

\section{Muon collider basics}
\label{sec:MC}

In this work we will assume the muon collider specifications laid out in Ref. \cite{Delahaye2013}. Specifically, we assume energies from $1.0 -6\, \tev$, a beam resolution of $\sim 0.1\%$ and a integrated luminosity of at least $500 \ifb$ \footnote{Nominal luminosities are: $\lum{1.25}{34}$ ($\sqrt{s} = 1.5\, \tev$),  $\lum{4.4}{34}$ ($\sqrt{s} = 3.0\, \tev$) \cite{Delahaye2013} and $\lum{1.2}{35}$ ($\sqrt{s} = 6.0\, \tev$) \cite{rpalmer2012}.}

The muon collider is already known to have unique capabilities as a SM-like Higgs factory~\cite{Barger:1996jm}; including  the measurement of Higgs mass to $\delta M(h^0) = 0.06\, \mev$, a direct width measurement to $\delta \Gamma = 0.15\,\mev$ and sensitivity to second generation couplings through the measurement of $Br(\mu^+\mu^-)\times Br(WW^*)$ to two percent\,\cite{Han:2012rb}.  Studies are ongoing for the accuracy of other Higgs couplings (see eg. \cite{Conway:2013lca}).  Because the SM Higgs width so small, $\Gamma_h/m_h = 3.4\times10^{-4}$, the requirements on a muon collider ($R=\delta E/E \sim 3 \times 10^{-5}$ with luminosity of approximately $10^{32}$ cm$^2$ sec $^{-1}$~\cite{Neuffer2013}) are very demanding~\cite{MChiggs2013}.  

\subsection{Resonant production}
\label{ssec:coll}

 To see why a muon collider is so well suited to $H/A$ production, we need to understand how the production rate is affected by various physical scales. The cross section for resonant $H/A$ production with subsequent decay into a final state $X$ is given by: 
\begin{align}
\label{eq:BW}
  \makebox[2em][l]{$\sigma(\mu^+\mu^- \ra H/A \ra X)$} & \nonumber \\
  & ~~~= \frac{4\pi\, \Gamma^2_{H/A}\, B_X}{(\hat s - m^2_{H/A})^2 + \Gamma^2_{H/A}\, m^2_{H/A}} 
\end{align}
where $B_X = Br (H/A \ra \mu^+\mu^-) Br (H/A \ra X)$ and $\sqrt{\hat s}$ is the center of mass energy of the collider. This parton-level cross section must be convolved with the beam energy resolution, taken as a Gaussian with variance $\Delta = R\,\sqrt s/\sqrt{2}$, where $\sqrt s$ is the nominal beam energy~\cite{Han:2012rb}.
\begin{equation}
\sigma_{eff}(s) = \int d\sqrt{\hat s}\, \sigma(\hat s)\times \text{Gauss}(\sqrt{s}, \Delta)
\end{equation}
When $\Delta \ll \Gamma_{H/A}$, the beam-smearing function collapses to a delta function. On resonance ($\sqrt s = m_{H/A}$), the cross-section then becomes a constant times the product of branching ratios
\begin{equation}
\sigma(\mu^+\mu^- \ra H/A \ra X) = \Big( \frac{4\pi}{m^2_{H/A}} \Big)\, B_X.
\end{equation}

In the opposite limit, $\Delta \gg \Gamma_{H/A}$, the rate is
\begin{equation}
\sigma(\mu^+\mu^- \ra H/A \ra X)  \sim \frac{B_X}{m^2_{H/A}}\Big(\frac{\Gamma_{H/A}}{\Delta} \Big),
\end{equation}
 and is suppressed relative to the previous limit by $\sim \Gamma_{H/A}/(4\pi\Delta)$.

The highest rate will therefore be achieved for particles that are wide compared to the beam resolution $\Delta \gg \Gamma_{H/A}$, yet narrow enough that  $Br(H/A \ra \mu^+\mu^-)$ is not infinitesimal. To get a rough idea for how large $Br(H/A \ra \mu^+\mu^-)$ needs to be, we can plug in some numbers  assuming $\Delta \gg \Gamma_{H/A}$ and $Br(H/A \ra X) \sim O(1)$: 
\vspace{0.1in}
\begin{eqnarray}
 && \makebox[0pt][l]{$\text{Events}/\text{year} = 1.54\times 10^5  $} \\ \label{eq:guessnum}
 && \times \Big( \frac{\mathcal L}{10^{34}\, \text{cm}^{-2}\,\text s^{-1}} \Big) 
  \Big( \frac{1\, \tev}{m_{H/A}} \Big)^2\Big(\frac{BR(H/A \ra \mu^+\mu^-)}{10^{-4}}\Big) \nonumber
\end{eqnarray}

\subsection{$H/A$ widths:}
\label{ssec:widths}

While there is still much to be learned about the couplings and properties of the $125\, \gev$ Higgs from future LHC analyses, the couplings of the Higgs boson to $W^{\pm}$ and $Z^0$ bosons are already limited to be close to their SM values~\cite{LHChiggs,Aaltonen:2013kxa}. Within a 2HDM, the $hVV$ couplings are set by $\sin(\beta-\alpha)$ where $\tan{\beta}$ is the ratio of vacuum expectation values and $\alpha$ is the measure of how much the two CP-even, neutral components of the two Higgses mix to form the mass eigenstates. Nearly SM-like $hVV$ couplings implies $\sin(\beta-\alpha) \sim 1$, the so-called `alignment limit'~\cite{Craig:2013hca}. With the current data, the deviation from the alignment limit is restricted to be $|\sin(\beta-\alpha) - 1| \lesssim 10\%$~\cite{Craig:2013hca} in a Type-II 2HDM\footnote{This estimate is based on tree level couplings and neglects new-physics contributions to production and decay. In Type-I 2DHM, Ref.~\cite{Craig:2013hca} show larger deviation from the alignment limit is allowed, especially at large, up to $O$(50\%) for large $\tan{\beta}.$.}

The coupling of the heavy Higgses to $W^+W^-$ and $Z^0Z^0$ are governed by the complementary function $\cos(\beta - \alpha)$. As the $125\,\gev$ Higgs couplings approach their SM values, the heavy Higgses decouple from $W^+W^-/Z^0Z^0$.  For sizable $HVV$ couplings, the decay width to $VV$ grows rapidly with Higgs mass $\Gamma/m_H \sim m^2_H$. For decoupled $H/A$, this growth is absent; the remaining (tree level) decay channels are (predominantly) fermionic, which lead to $\Gamma/m_H \sim \text{const}\times \frac{m^2_f/v^2}{16\pi^2}$. 
From this expression, one may worry that the combination of heavy fermions accompanied by large, i.e $\tan{\beta}$-enhanced couplings could generate a large width. However, this combination does not occur in supersymmetry or Type II 2HDM. The couplings of the top quark to $H/A$ are suppressed by $\tan{\beta}$, while the couplings of $H/A$ to $b/\tau$ are enhanced by $\tan{\beta}$ but come with the price of the small  bottom/tau mass\footnote{For Type-I 2HDM {\em all} SM fermion decay modes of $H/A$ are suppressed by $\tan{\beta}$}.
Finally, there is a $Z^{\mu}(H\partial_{\mu} A)$ interaction that could conceivably lead to a $m_H$-enhanced partial width, however this mode is typically strongly suppressed by phase space since the $H/A$ are often nearly degenerate.

Extra light matter that interacts with $H/A$ will also contribute to the width, so there is no model-independent way to guarantee that $H/A$ remain narrow, even in the alignment limit. However, in scenarios where the extra matter is only weakly coupled to $H/A$, such as supersymmetry, the $H/A$ width remains low. 

While the fact that the $hVV$ couplings are close to their SM values allows us to make fairly general statements about the $H/A$ width, we know much less about the $H/A$ mass. Within supersymmetry, the Higgs potential simplifies (compared to general 2HDM form) and deviations from the alignment limit can be expressed in terms of the $m_A$:
\begin{equation}
\cos(\beta - \alpha)|_{\text{align}} = \frac{m^2_Z\,\sin{4\beta}}{2\,m^2_A}.
\end{equation}
Because of the quadratic dependence on $m_A$, limits on $\sin(\beta- \alpha)$ need to improve significantly before indirect limits on $m_A$ limits are pushed much above the weak scale.  

While the widths we are discussing are low, a $\Gamma_H/m_H \sim \frac 1 {16\pi^2}$ is still an order of magnitude larger than the nominal muon collider beam-energy resolution. 

\section{Benchmark Scenarios}
\label{sec:bench}

We illustrate the general picture with specific supersymmetry benchmark examples for which complete spectra are specified.  For ease of comparison, we will use some of the supersymmetry spectra proposed in Ref.~\cite{snowmass} as benchmarks for Linear Collider studies.

A comparison of the cross section for resonant $H/A$ production with other Higgs production processes is shown below in Fig.~\ref{fig:rates}. The top panel of Fig ~\ref{fig:rates} compares the cross sections for s-channel resonances $H/A$ for a number of different supersymmetry benchmark models.  In all these cases, the s-channel rates dominate the cross sections for associated production of the light SM-like Higgs.   The bottom panel of Fig.~\ref{fig:rates} compares the s-channel for the Natural Supersymmetry benchmark with other production modes for these heavy Higgses available to both a muon collider and a linear electron collider.   The resonant production  available in a muon collider is over two orders of magnitude larger than the other processes.
\begin{figure}[h!]
\centering
\includegraphics[width=0.49\textwidth, height=2.5in]{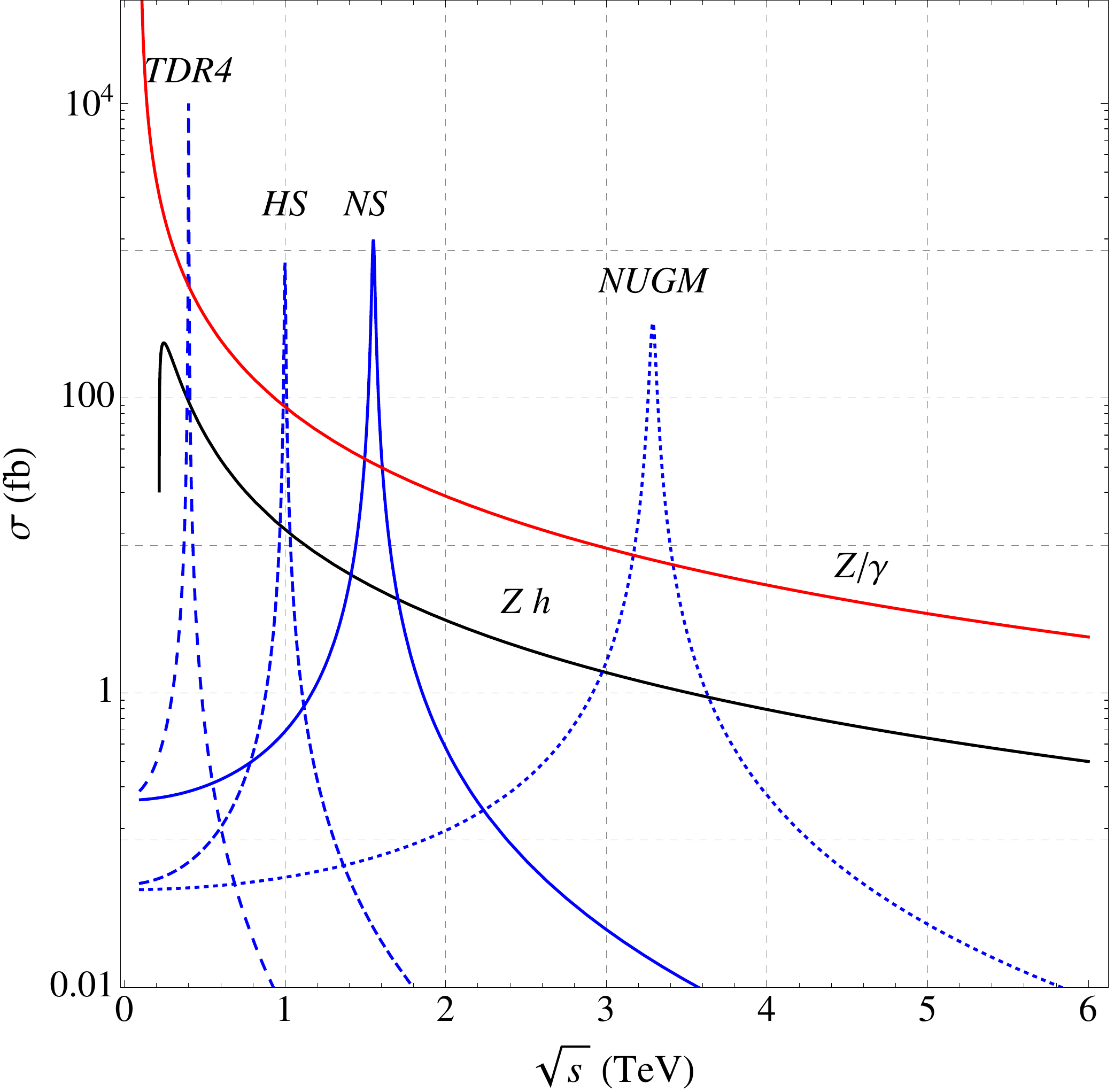} 
\includegraphics[width=0.49\textwidth, height=2.5in]{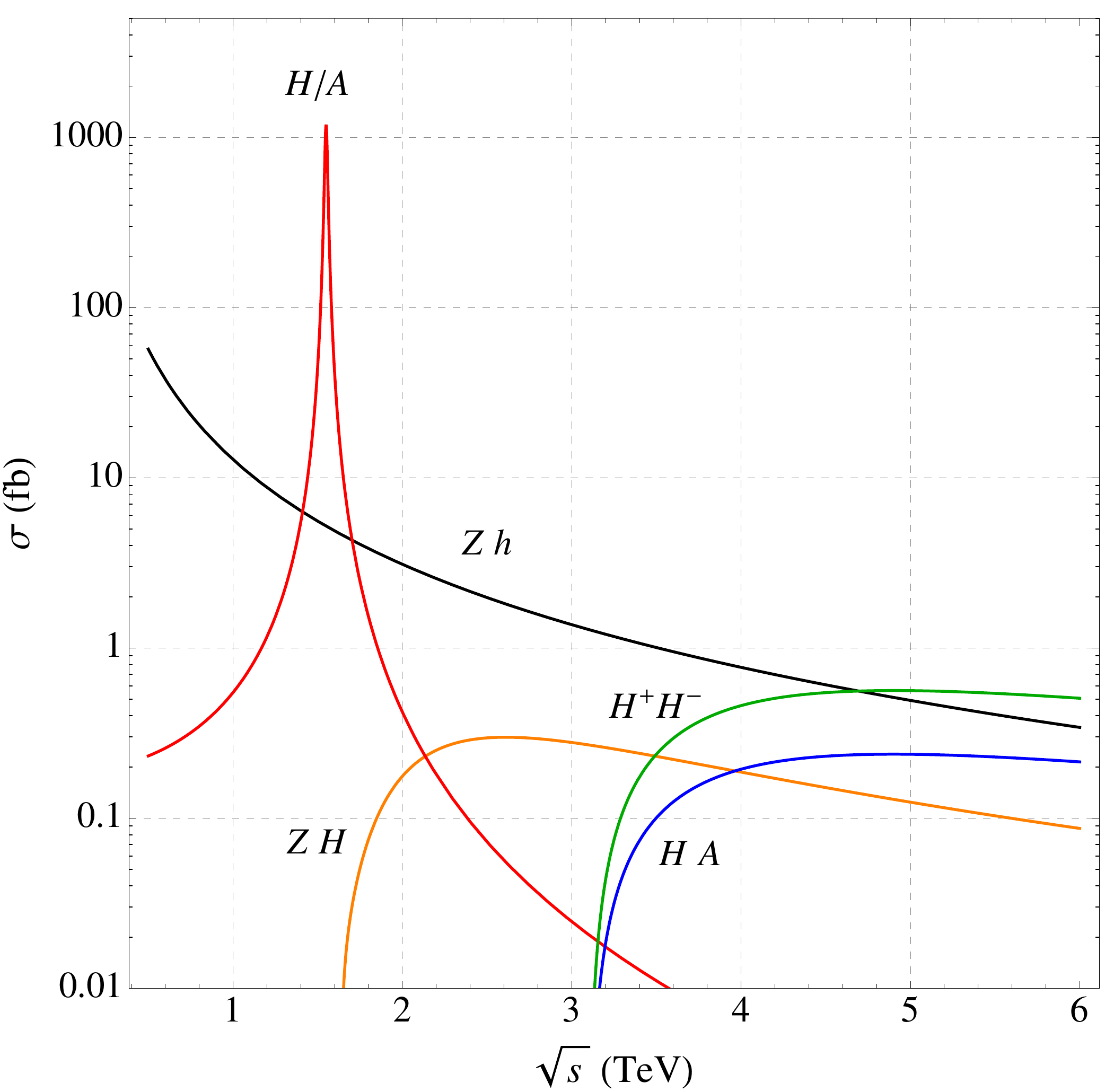}
\caption{Top panel: comparison of resonant $H/A$ production in several benchmark supersymmetry scenarios~\cite{snowmass} with $Z^0h$ and $\gamma^*/Z^0$ production. The models are: HS = Hidden Supersymmetry, NS = Natural Supersymmetry, NUGM = non-universal Higgs mass, and TDR4 = light-slepton, stau NLSP model. For the complete spectra in these scenarios, see Ref~\cite{snowmass}. Bottom: Comparison of $H/A$ production in the Natural Supersymmetry model with $Z^0h$, $Z^0H$ and heavy Higgs pair production. In both plots $H/A$ production is the sum of $\mu^+\mu^- \ra H$ and $\mu^+\mu^- \ra A$ as the states are nearly degenerate.}
\label{fig:rates}
\end{figure}

\section{Natural Supersymmetry Example}
\label{sec:NS}

In order to study the opportunities of the muon collider as a $H/A$ factory in detail, we focus here only one benchmark of Ref.~\cite{snowmass} the Natural Supersymmetry model. The masses and principal decay modes of the $H/A$ in this model are given in Table \ref{tab:HAprop}. 
\begin{table}[htbp]
   \centering
   \caption{Properties of the  $H$ and $A$ states in the Natural Supersymmetry benchmark model \cite{snowmass}.  In addition to masses and total widths,  the  branching ratios for various decay modes are shown.}
      \begin{tabular}{@{} ccccc @{}}      
       & \multicolumn{2}{c}{$H$} &\multicolumn{2}{c}{$A$} \\[2mm]
       Mass & \multicolumn{2}{c}{$1.560\, \tev$}  & \multicolumn{2}{c}{$1.550\, \tev$} \\[1mm]
       Width & \multicolumn{2}{c}{$19.5\, \gev$}  & \multicolumn{2}{c}{$19.2\, \gev$} \\[1mm]
       & (Decay) & Br  & (Decay)  & Br   \\[2mm]
       & ($b\bar b$) & $0.64$ & ($b\bar b$) &   $0.65$  \\[1mm]
       & ($\tau^+ \tau^-$) & $8.3\times 10^{-2}$ & ($\tau^+ \tau^-$) & $8.3\times 10^{-3}$  \\[1mm]
       & ($s\bar s$)   & $3.9\times 10^{-4}$ &($s\bar s$) & $4.0\times 10^{-3}$  \\[1mm]
       & ($\mu^+ \mu^-$) & $2.9 \times 10^{-4}$ &  ($\mu^+ \mu^-$) & $2.9 \times 10^{-4}$  \\[1mm]
       & ($t \bar t$) & $6.6\times 10^{-3}$  & ($t \bar t$) & $7.2\times 10^{-3}$  \\[3mm]
   
       & ($g g$) & $1.4 \times 10^{-5}$  & ($g g$) & $6.1 \times 10^{-5}$ \\[1mm]
       & ($\gamma \gamma$) & $1.1\times 10^{-7}$  & ($\gamma \gamma$) & $3.8\times 10^{-9}$  \\[1mm]
       & ($Z^0 Z^0$) & $2.6 \times 10^{-5}$ &  ($Z^0 \gamma$) & $4.3 \times 10^{-8}$   \\[1mm]
       & ($h^0 h^0$) & $4.4 \times 10^{-5}$ & &   \\[1mm]   
       & ($W^+W^-$) & $5.3 \times 10^{-5}$ & &  \\[3mm]
            
       & ($\tilde{\tau}_1^\pm \tilde{\tau}_2^{\mp}$) & $9.2 \times 10^{-3}$ & ($\tilde{\tau}_1^\pm \tilde{\tau}_2^{\mp}$) & $9.5 \times 10^{-3}$ \\[1mm]\
       & ($\tilde{t}_1 \tilde{t}_1^*$) & $3.1 \times 10^{-3}$ & ($\tilde{t}_1 \tilde{t}_2^*$)  & $1.1\times 10^{-3}$ \\[3mm]
       & ($\chi_1^0 \chi_1^0$) & $2.6 \times 10^{-3}$ & ($\chi_1^0 \chi_1^0$) & $3.2 \times 10^{-3}$ \\[1mm]
       & ($\chi_2^0 \chi_2^0$) & $1.3 \times 10^{-3}$ & ($\chi_2^0 \chi_2^0$) & $1.1 \times 10^{-3}$ \\[1mm]
       & ($\chi_1^0 \chi_3^0$) & $2.8 \times 10^{-2}$ & ($\chi_1^0 \chi_3^0$) & $3.9 \times 10^{-2}$ \\[1mm]
       & ($\chi_1^0 \chi_4^0$) & $1.7 \times 10^{-2}$ & ($\chi_1^0 \chi_4^0$) & $4.0 \times 10^{-2}$ \\[1mm]
       & ($\chi_2^0 \chi_3^0$) & $3.8 \times 10^{-2}$ & ($\chi_2^0 \chi_3^0$) & $2.7 \times 10^{-2}$ \\[1mm]
       & ($\chi_2^0 \chi_4^0$) & $4.0 \times 10^{-2}$ & ($\chi_2^0 \chi_4^0$) & $1.5 \times 10^{-2}$ \\[1mm]
       & ($\chi_1^\pm \chi_2^{\mp}$) & $5.7 \times 10^{-2}$  & ($\chi_1^\pm \chi_2^{\mp}$) & $6.0 \times 10^{-2}$ 
     \end{tabular}
     \label{tab:HAprop}
\end{table}

First, we consider cross section for the largest decay mode of the $H/A$, i.e. $b\bar b$.  Since a muon collider requires shielding in the forward and backward cones,  we make fiducial cuts at $10^o$ about the beam axis.  In Fig. \ref{fig:ns_bbspect} the $b \bar b$ cross section is shown for a scan from $\sqrt{s} = 1450-1650 \, \gev$ in 100 steps of $2 \, \gev$ with a luminosity of $5.0 \ifb$ per step. The cross section at a given nominal luminosity is calculated using PYTHIA6~\cite{Sjostrand:2006za} with modifications to include gaussian beam energy smearing with a resolution parameter $R=0.001$.  As can be seen from the top panel of  Fig.~\ref{fig:ns_bbspect}, the peak signal is more than an order of magnitude larger than the background.  

\begin{figure}[h!]
\centering
\includegraphics[width=0.45\textwidth]{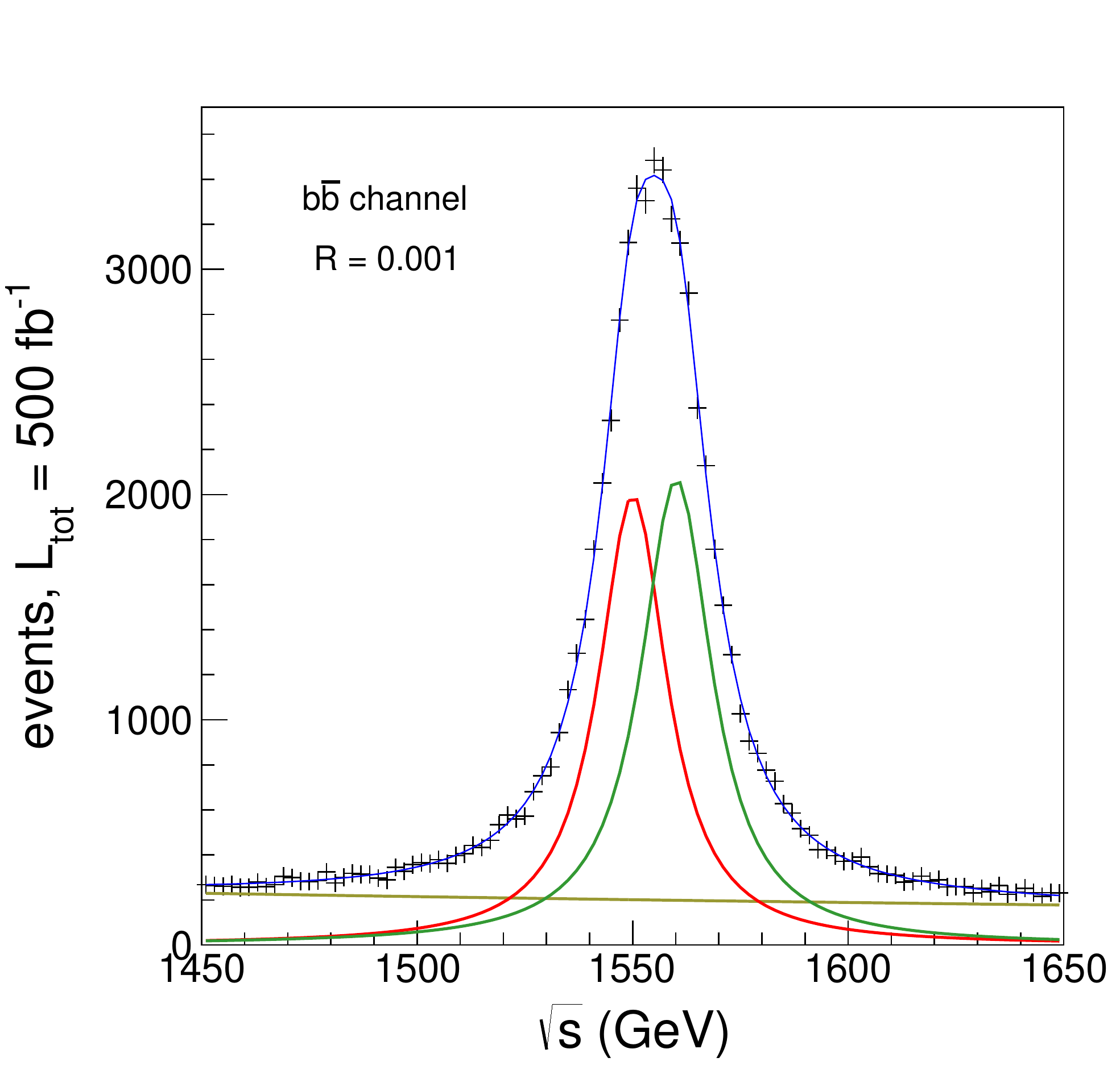} \\
\includegraphics[width=0.45\textwidth]{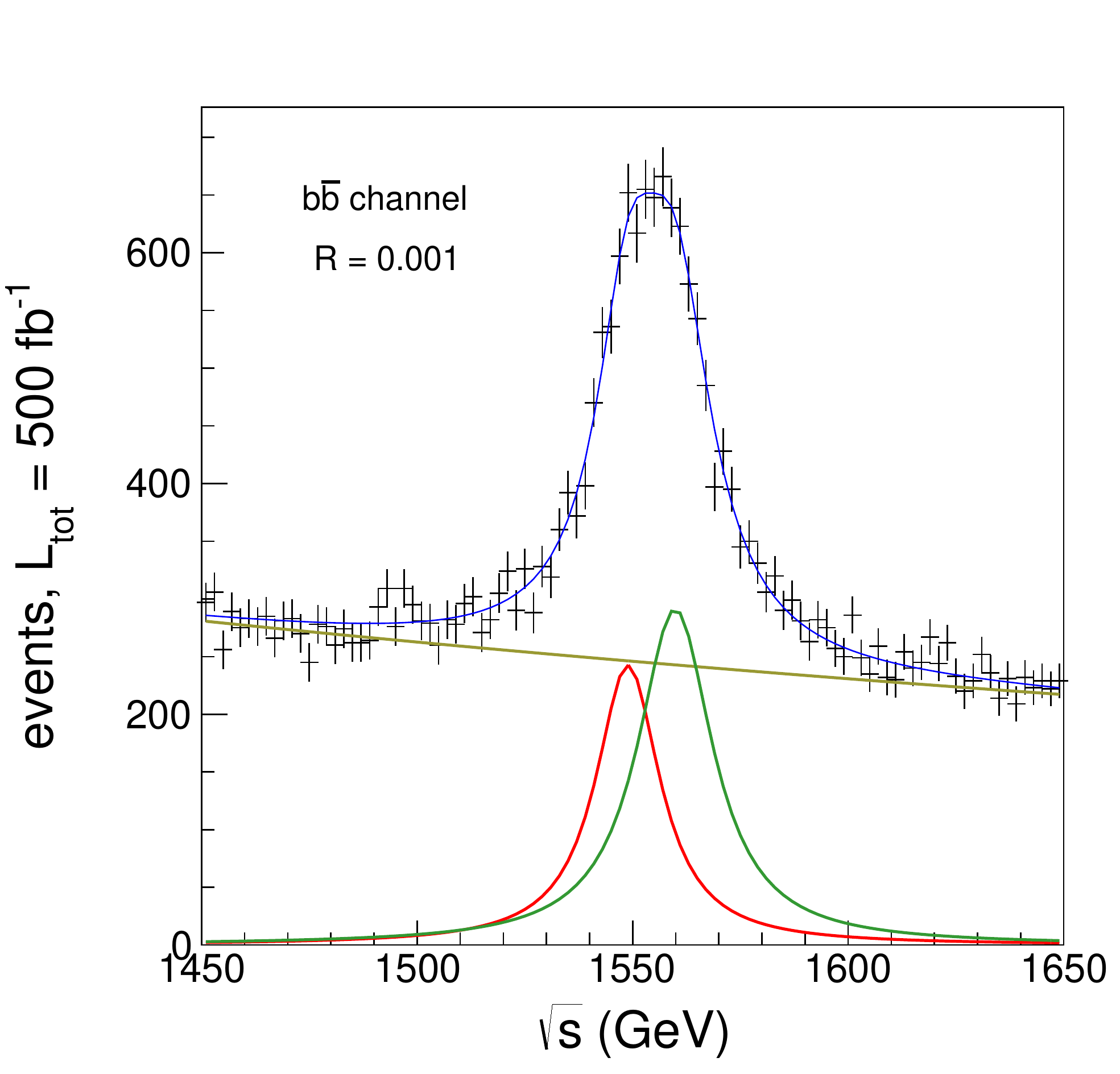} 
\caption{Pseudo-data (in black) along with the fit results in the $\bar b b$ (top) and $\tau^+\tau^-$ (bottom) channels. The two Breit-Wigner components ($A$ in green, $H$ in red) along with the background component (yellow) are also shown. In each bin, the expected number of events - the PYTHIA cross section times $5\,\ifb$ was allowed to fluctuate according to Poisson statistics.}
\label{fig:ns_bbspect}
\end{figure}

We use this channel to study the ability of extracting separate information about the two nearby resonances.  We fit the cross section in this region by a sum of background,  $\sigma_B$ given by:
\begin{equation}
  \sigma_B(\sqrt{s})= c_1 \frac{(m_H m_A)}{s}
\end{equation}
and one or two Breit-Wigner's (Eq. \ref{eq:BW})
for the signal contributions.  

The resulting fits are shown in Table \ref{tab:HAfit}.  A single Breit-Wigner is completely ruled 
out while the two resonance fit provides an excellent description of the total cross section 
and allows an accurate determination of the individual masses, widths and $B_{b\bar b}$ branching ratios of the $A$ and $H$.

\begin{table}[htbp]
   \centering
   \topcaption{Fit of the $H/A$ region to background  plus  Breit-WIgner resonances. 
Both a single and two resonance fits are shown.  General form of the background fit is $\sigma_B(\sqrt{s}) =  c_1 (1.555)^2/{s ({\rm~in~} \tev^2)}$. The values of the best fit 
for one or two Breit-Wigner resonances are given.}
   \begin{tabular}{@{} ccc @{}}      
\multicolumn{3}{c}{One Resonance} \\[1mm]
${\rm Mass} (\gev)$ & $\Gamma (\gev)$ & $\sigma_{\rm peak}$ (pb) \\[1mm]
   $1555 \pm 0.1 \, \gev$ & $24.2 \pm 0.2$ & $1.107 \pm 0.0076$ \\[1mm]
   $\chi^2/{\rm ndf} = 363/96$ & &$c_1 = 0.0354 \pm 0.0006$  \\[3mm]
\multicolumn{3}{c}{Two Resonances} \\[1mm]
 ${\rm Mass} (\gev)$ & $\Gamma (\gev)$ &  $\sigma_{\rm peak}$ (pb) \\[1mm]
 $1550 \pm 0.5 \, \gev$ & $19.3 \pm 0.7$ & $0.6274 \pm 0.0574$ \\
 $1560 \pm 0.5 \, \gev$ & $20.0 \pm 0.7$ & $0.6498\pm 0.0568$ \\[1mm]
  $\chi^2/{\rm ndf} = 90.1/93$ & & $c_1 = 0.040 \pm  0.0006 $ \\
   \end{tabular}
   \label{tab:HAfit}
\end{table}

\section{$H/A$ factory}
\label{sec:factory}
   
In the previous section, we investigated  the principal decay mode of the $H/A$ resonances,  the $b\bar b$ channel.  We have determined the masses, total widths and branching ratio
$Br(\mu^+\mu^-)\times Br(b\bar b)$ for both the $H$ and $A$.  Now we consider other decay modes. 

\subsection{The $\tau^+\tau^-$ decays}

The $\tau$ pair branching fractions are typically large ($\sim$10\%) and so we have high statistics  for this mode as well.  The signal cross section to the final state $\tau^+\tau^-$ is shown in Fig. \ref{fig:ns_bbspect}.  The signal to background ratio is $S/B \cong 1.5$.  So we can fit the individual states with the same form of two Breit-Wigner resonances and a background as before. 
 Here we use the masses and widths already determined from the $b\bar b$ channel.  So we have 4 parameters (two for background fits and a peak cross section for each of the two resonances).  This allows the extraction of relative branching fraction $R_{\tau+\tau^-} = Br(\tau^+\tau^-)/Br(b\bar b)$ for each state.  We obtain $R_{\tau+\tau^-} (A) = 0.141\pm0.014$ and   $R_{\tau+\tau^-} (H) = 0.121\pm 0.013$.
 
Furthermore, we may be able to use the large rate for the $\tau^+\tau^-$ decay modes to determine the CP properties of the $H$ and $A$. 
When both $\tau^{\pm}$ decay hadronically, one can determine the CP properties~\cite{Worek:2003zp}. The practicality of this in the real muon collider environment and with
a feasible detector needs to be investigated. The use of  rare decay modes that are not common to both $H$ and $A$: $H\ra Z^0Z^0, W^+W^-, hh$, and $A \ra Z^0\,h$ seems more problematic.  Within the benchmark scenarios, the branching fractions to these modes are $\lesssim 10^{-3}$. After paying the price of additional $V/h$ branching fractions to clean final states (to avoid mass overlap), we are left with only a few events per year according to Eq. (\ref{eq:guessnum}).

\subsection{Decays to neutralinos and charginos}

If there are additional light states that interact with the $H/A$, rare decays of $H/A$ offer an additional, often complementary production mechanism. Within the context of supersymmetry, the elctroweakinos (bino, winos, Higgsinos) and sleptons are two such examples. The $H/A$ branching fractions to electroweakinos and sleptons are each $O$(few \%) (see Table \ref{tab:HAprop}),  leading to an effective cross section of $O(70\, \fb)$. As the Lorentz structure of Yukawa and gauge interactions are different, sleptons/inos produced from $H/A$ decay will have a different handedness structure compared to  events produced via $\gamma^*/Z^0$. Thus, by combining both production mechanisms, we become sensitive to a wider set supersymmetry parameters. \\

Because the resonance production of $H$ or $A$ is only through the scalar channel, any polarization of the initial muon beams does not 
change the relative production of the electroweakino final states.  Thus for these processes the small $(\sim15\%)$ polarization of initial muon beams
will not adversely affect the physics sensitivity.

As the superpartner mass is increased, $H/A$ decay quickly becomes the dominant production mechanism, and, at some point, a muon collider $H/A$ factory is the only feasible way to study certain parts of the spectrum. One example is the mixed chargino state $\chi^{\pm}_2\chi^{\mp}_1$ within the Natural Supersymmetry benchmark; as $m_{\chi_2} \sim \tev$, EW production of this final state is tiny $\ll \text{fb}$, both at the LHC and a $\sqrt s = m_{H/A} \sim 1.5\,\tev$ muon collider. However, $Br(H/A \ra \chi^{\pm}_2\chi^{\mp}_1$) is still $O$(few \%), so thousands of $\chi^{\pm}_2\chi^{\mp}_1$ pairs would be produced per year from $H/A$ decay\footnote{Access to heavier electroweakinos is especially important in scenarios, such as the Natural Supersymmetry setup, where the light inos are compressed $m_{\chi^{\pm}_1} \sim m_{\chi^{0}_1}$ and are therefore difficult to probe at the LHC}.

The invisible decay modes of $H/A$ are expected to be very small (see Table \ref{tab:HAprop}, but could be probed by running at $\sqrt s $ slightly above the $m_{H/A}$ resonances.  The incoming muons can return to the resonant $\sqrt s$ by emitting a photon. This photon can help tag otherwise invisible decays\footnote{This approach is not specific to neutralinos. It can be used for any $H/A \ra $ invisible decays, such as those arising from dark matter Higgs portal~\cite{Silveira:1985rk,McDonald:1993ex,Burgess:2000yq,Bai:2012nv} interactions}. The feasibility of this method needs further study.

\section{Discussion}
\label{sec:disc}

  The muon collider is a ideal $H/A$ factory for masses well into the multi-TeV range.  The s-channel production of the heavy spin-zero $H$ and $A$ at a muon collider provides 
 a unique opportunity to study the properties of the states.  We have shown the masses and widths
 of the individual states can be disentangled with high accuracy even for nearly degenerate
 states.  Furthermore, the tau pair  decay's allow the measurement of the individual CP of these two states.  Finally, the decay products include large rates for supersymmetric pairs, mostly neutrinos and charginos from a well defined initial state.  This eliminates the need for polarization of the initial muon beams for these studies.
 
Two important issues need to be addressed in order to realize this potential for $H/A$.  First,
 these result have to be validated with a realistic detector simulation and the shielding at the machine-detector interface.  The shielding is required to suppress the  substantial backgrounds arising from muon decays in the beams that will flood the detector with (mainly low energy) photons, neutrons, and other particles.  Second, we need to to find the $H/A$ resonance.  In the context of supersymmetric models, the LHC direct observation of these heavy scalars will be limited to $m_{H/A} \sim 1\, \tev$ depending on $\tan{\beta}$. If the $H/A$ are
more massive, we will need theoretical guidance from the observation of other light sparticles  or small deviations in the SM-like Higgs branching fractions.  In the worst case, the muon collider 
at high energy (e.g. $6 \, \tev$) could possibly be used to both study new physics and search for any $H/A$
resonance below this energy using the small associated production or initial state radiation cross sections.

We emphasize that while we have performed detailed analysis on a particular supersymmetric 2DHM, our main results are not sensitive to this choice. Measurements of the $125\,\gev$ Higgs imply the heavy Higgses -- if they exist -- are generically narrow. Provided $\Gamma_{H/A}$ is larger than the beam-energy resolution, narrow, heavy Higgses can be produced abundantly at a muon collider, regardless of whether they are part of a supersymmetry, or even a Type-II 2HDM.  The possible final states and exact branching fractions will, however, depend on the model.\\

\section*{Acknowledgments}
We thank N. Craig for helpful discussions.
This work was supported by Fermilab operated by Fermi Research Alliance, LLC,
U.S.~Department of Energy Contract~DE-AC02-07CH11359 (EE).

\end{document}